\documentclass[conference]{IEEEtran}
\IEEEoverridecommandlockouts

\usepackage{mdframed}     % For creating the framed box with rounded corners
\usepackage{lipsum}       % Optional: just for dummy text around the figure

\usepackage{amsmath}
\usepackage{amsthm}
\usepackage{times}
\usepackage{breqn}

\usepackage[normalem]{ulem}

\usepackage{algorithm}
\usepackage{algpseudocode}
\algrenewcommand\algorithmicrequire{\textbf{Input:}}
\algrenewcommand\algorithmicensure{\textbf{Output:}}

\usepackage{epsfig}
\usepackage{epstopdf}
\usepackage{subfigure}
\usepackage{multirow}
\usepackage{comment}
\usepackage{listings}
\usepackage{xcolor}
\usepackage{soul}
\usepackage{xspace}
\usepackage{url}
\usepackage{rotating}

\usepackage{booktabs}     % For professional-looking horizontal lines (\toprule, \midrule, \bottomrule)
\usepackage{array}        % For advanced column formatting (e.g., p{width}, >{\centering\arraybackslash})
\usepackage{makecell}     % For creating multi-line cells easily (\makecell)
\usepackage{multirow}               % (only needed if you plan to add multi‑row cells later)
\usepackage{adjustbox}              % lets the table auto‑shrink to text width

\renewcommand{\arraystretch}{1.25}   % additional row‐height scaling

\usepackage{cite}
\usepackage{amsmath,amssymb,amsfonts}
\usepackage{graphicx}
\usepackage{textcomp}
\usepackage{xcolor}
\def\BibTeX{{\rm B\kern-.05em{\sc i\kern-.025em b}\kern-.08em
    T\kern-.1667em\lower.7ex\hbox{E}\kern-.125emX}}

\graphicspath{{figures/}}

\makeatletter
\newcommand{\linebreakand}{%
  \end{@IEEEauthorhalign}
  \hfill\mbox{}\par
  \mbox{}\hfill\begin{@IEEEauthorhalign}
}
\makeatother

\usepackage{listings}
\usepackage{xcolor}

\lstdefinestyle{mystyle}{
    backgroundcolor=\color{gray!10},   
    commentstyle=\color{green!50!black}\itshape,
    keywordstyle=\color{blue}\bfseries,
    numberstyle=\tiny\color{gray},
    stringstyle=\color{orange},
    basicstyle=\ttfamily\footnotesize,
    breakatwhitespace=false,         
    breaklines=true,                 
    captionpos=b,                    
    keepspaces=true,                 
    numbers=left,                    
    numbersep=5pt,                  
    showspaces=false,                
    showstringspaces=false,
    showtabs=false,                  
    tabsize=2
}

\begin{document}

\title{Evaluating Efficiency and Novelty of LLM-Generated Code for Graph Analysis}
% 
%An empirical study on evaluating the efficiency and novelty of Large Language Model's generated code in C

% \author{
%     \IEEEauthorblockN{David A. Bader}
%     \IEEEauthorblockA{
%        \textit{Department of Data Science} \\
%         \textit{New Jersey Institute of Technology}\\
%         Newark, New Jersey, USA\\
%         bader@njit.edu
%     }
%     \and
%     \IEEEauthorblockN{Atieh Barati Nia}
%     \IEEEauthorblockA{
%         \textit{Department of Data Science} \\
%         \textit{New Jersey Institute of Technology}\\
%         Newark, New Jersey, USA\\
%         ab2763@njit.edu
%     }
% }

\author{
  \IEEEauthorblockN{Atieh Barati Nia,
                    Mohammad Dindoost,
                    David A. Bader}
  \IEEEauthorblockA{Department of Data Science, New Jersey Institute of Technology, Newark, NJ, USA\\
  \texttt{\{ab2763, md724, bader\}@njit.edu}}
  \thanks{This work was supported in part by NSF grants CCF-2109988 and OAC-2402560, and CCF-2453324.
}
}

\maketitle

\begingroup\renewcommand\thefootnote{$^*$}

\endgroup
%% article.
\begin{abstract}
% \textcolor{red}{Add a concise summary of the research motivation, methods, key findings (e.g., Claude 3.7 Sonnet Extended outperformed human-written code), and contribution to literature.}

% \textcolor{red}{I wrote the first version}

% As Large Language Models (LLMs) become increasingly integrated into software development, their efficiency in graph algorithms remains unexplored, particularly for generating efficient low-level code in C. This paper presents a novel evaluation of state-of-the-art LLMs on the triangle counting in large graphs. Using a reproducible benchmarking framework and executing all tests on NJIT’s Wulver HPC cluster, we assess seven prominent models, including Claude 3.7 Sonnet Extended and OpenAI o3-mini-high, for correctness and runtime efficiency on RMAT graphs up to RMAT-18 scale. Our results show that while some models generate incorrect or inefficient code, others match or surpass optimized human-written implementations. In particular, Claude 3.7 Sonnet Extended outperforms all other models in runtime performance without introducing fundamentally new algorithms, demonstrating LLMs' strength in optimization over innovation. This study contributes the first C-level performance benchmark for LLM-generated code and provides a foundation for future research in GPU support, parallel code generation, and robustness in HPC contexts.

Large Language Models (LLMs) are increasingly used to automate software development, yet most prior evaluations focus on functional correctness or high-level languages such as Python. As one of the first systematic explorations of LLM-assisted software performance engineering, we present a comprehensive study of LLMs' ability to generate efficient C implementations of graph-analysis routines—code that must satisfy stringent runtime and memory constraints. This emerging field of LLM-assisted algorithm engineering holds significant promise, as these models may possess the capability to design novel approaches that improve existing algorithms and their implementations. Eight state-of-the-art models (OpenAI ChatGPT o3 and o4-mini-high, Anthropic Claude 4 Sonnet and Sonnet Extended, Google Gemini 2.5 Flash and Pro, xAI Grok 3-Think, and DeepSeek DeepThink R1) are benchmarked using two distinct approaches. The first approach evaluates the ability of LLMs to generate algorithms that outperform existing benchmarks. The second approach assesses their capability to generate graph algorithms for integration into performance-critical systems. The results show that Claude Sonnet 4 Extended achieves superior performance in ready-to-use code generation and efficiency, outperforming human-written baselines in triangle counting. Although our findings demonstrate that contemporary LLMs excel in optimizing and integrating established algorithms, the potential for these models to eventually invent transformative algorithmic techniques represents a compelling frontier for future research. We provide prompts, generated code, and measurement scripts to promote reproducible research in this rapidly evolving domain.  All of the source code is available on GitHub at https://github.com/Bader-Research/LLM-triangle-counting/.
\end{abstract}


\section{Introduction}
% \textcolor{green}{Emphasize the novelty of studying C code efficiency, especially since Python has been more commonly studied.}

% \textcolor{green}{To the best of our knowledge, this is the first work that evaluates the code generation of models by giving files and complicated problem defining. and large data 
% normalized time usage
% normalized memory usage
% bold the least time run}

% \textcolor{green}{... in selecting the most suitable LLMs for their real-time applications. Consider briefly highlighting why triangle counting is a representative and challenging problem. In General we should have a reason for selecting C and Algorithms
% }

The rapid advancement of Artificial Intelligence (AI), particularly large language models (LLMs), has significantly impacted both everyday life and scientific workflows, accelerating research and development across disciplines. LLMs have shown significant promise in improving learning, automating tasks, and boosting productivity in a variety of domains. A recent survey by Chang \emph{et al.}\cite{chang2024survey} highlights a comprehensive review of evaluation efforts related to LLM applications in diverse areas, including reasoning, medical applications, ethics, and education. One particularly impactful application of LLMs is to help programmers and computer scientists by streamlining software development and minimizing human error. As a result, considerable research has focused on evaluating the correctness of code generated by these models \cite{chen2022codet, liu2023your, tong2024codejudge, nguyen2022empirical} and some research's focus is on correctness and debugging\cite{dinh2023large, tambon2025bugs, jesse2023large}. Benchmarks such as HumanEval \cite{chen2021evaluating} and MBPP \cite{austin2021program} have been widely used to evaluate the functional correctness of LLM-generated code through the completion of unit tests. Although these benchmarks assess whether the code works as intended, they do not account for performance, which is the focus of our study. 

Although Python dominates recent research, computer scientists still depend on lower-level languages such as C and C++ to achieve peak performance. Assessing the efficiency of LLM generated C code is therefore vital to judging the ability of these models to produce optimized implementations. However, to our knowledge, no study has systematically examined the performance of LLM-generated code in C, a language central to high‑performance and scientific computing.

Moreover, prompting an LLM to generate an algorithm without explicit contextual constraints may raise issues of construct validity because we cannot know whether the model is simply repeating code it saw during training. To address this concern, we adopt a more rigorous evaluation strategy, the first to our knowledge that places the models inside an existing codebase, by supplying concrete source files and asking the LLMs to produce algorithms that integrate with them. This targeted setup sharply reduces the chance of pre‑training contamination and yields a clearer measure of each model’s genuine coding ability.

This paper offers the first systematic evaluation of cutting-edge LLMs in producing high-performance C implementations for graph algorithms. We evaluated the generated code along four axes: compilability, functional correctness, execution time, and memory footprint. We anticipate that our findings will not only guide LLM developers in enhancing their models but also assist those computer and computational scientists, who implement application codes or graph analytics, in selecting the most suitable LLMs for their real-world applications.

\subsection{Related work}

% \textcolor{green}{... in producing performant code. Consider synthesizing how prior Python-based benchmarks fall short for HPC and justify the focus on C more strongly.}

Recent research has increasingly focused on the efficiency of code generated by large language models (LLMs), moving beyond the traditional emphasis on correctness. Huang \emph{et al.}\cite{huang2024effibench} introduce EffiBench, a benchmark that evaluates the runtime and algorithmic complexity of Python code, revealing that syntactically correct code is not always efficient. Similarly, Qiu \emph{et al.}\cite{qiu2024efficient} propose a rigorous benchmark to jointly assess both the correctness and efficiency of the LLM-generated Python code. Licorish \emph{et al.}\cite{licorish2025comparing} compares the human-generated and LLM code in correctness, efficiency, and maintainability, concluding that current models still lack human performance. Du \emph{et al.}\cite{du2024mercury} presents Mercury, a broad efficiency benchmark for LLMs across multiple programming tasks, while Niu \emph{et al.}\cite{niu2024evaluating} and Liu \emph{et al.}\cite{liu2024evaluating} examine key challenges and proposed frameworks for measuring efficiency in generated Python code.

Although these studies yield valuable information for a popular high‑level language, C remains the workhorse of high‑performance computing (HPC) and system programming. Virtually all supercomputing kernels, MPI/OpenMP libraries, operating system components, and performance-critical embedded routines are still written in C (or C‑centric dialects such as C++, CUDA, and OpenCL). Unlike Python - whose interpreter and dynamic typing add layers of abstraction - C exposes explicit memory management and low‑level control over hardware, so a small inefficiency can translate into orders‑of‑magnitude slowdowns at HPC scale. Evaluating LLM output in C therefore probes a harder and more consequential problem: can these models generate code that meets the stringent runtime, memory, and power budgets demanded by real‑world analytics, exascale simulations, and kernel‑level software?

Other relevant efforts focus on improving the quality or efficiency of the code. Siddiq \emph{et al.}\cite{siddiq2024franc} introduces FRANC, a framework designed to improve the correctness and readability of LLM-generated Java and Python code. Yetistiren \emph{et al.}\cite{yeticstiren2023evaluating} empirically evaluate tools such as GitHub Copilot, Amazon CodeWhisperer, and ChatGPT, focusing on quality dimensions such as correctness and maintainability, but primarily using earlier LLMs. In contrast, our study evaluates the performance of code produced by state-of-the-art models.

Ye \emph{et al.}\cite{ye2025llm4effi} propose LLM4EFFI, a framework that guides LLMs to produce more efficient and correct Python code using prompt engineering and iterative feedback. Likewise, Waghjale \emph{et al.}\cite{waghjale2024ecco} introduces ECCO, a post-processing framework that optimizes the generated code for efficiency. Unlike these works, our study evaluates the inherent performance of the code generated by modern LLMs in C, without manual intervention or optimization, to reflect their true capabilities in producing performant code.

\section{Experiment Design}

In this section, we describe the data and source code of the program, the models we use for evaluation and comparison, and how we conduct experiments to investigate the performance of each LLM for assisting with software performance engineering tasks of optimizing the running time of graph algorithm code.

\subsection{Data}
\label{ss:data}

% \textcolor{red}{... to reduce the risk of using pre-trained data. You could clarify how this mitigates contamination, maybe mention the importance of context-aware generation.}

Bader \cite{bader2023fast} introduced a comprehensive framework in C to evaluate the performance of the existing corpus of human-generated triangle counting algorithms. Within this framework, he collected and implemented all known CPU-based triangle-counting algorithms for a sparse graph given in compressed sparse row (CSR) format, measured their execution times, and established a benchmark for comparison. This framework serves as a good input to the LLMs to incorporate their LLM-generated code and to reduce the risk of using pretrained data. 

\subsection{Models}

% \textcolor{green}{... that are publicly accessible and gained attraction... It would be helpful to indicate when each model was released, to show their currency.}

Since we want to study the ability of LLMs to assist programmers with software performance engineering, we selected the seven most frequently used frontier AI models available at the time of writing this paper and gained attraction because of their ability to generate source code. 

\textbf{ OpenAI ChatGPT o3 and ChatGPT o4-mini-high:} OpenAI's models are among the most well-known LLMs, mostly known as ChatGPT, having played a pivotal role in ushering in the modern era of generative AI. So, we included the evaluation of the latest available versions related to code implementation which were released on April 16, 2025. OpenAI o3 which uses advanced reasoning by spending more time thinking before responding, and OpenAI o4-mini-high which said to be great at coding and visual reasoning are selected to be the representative of OpenAI company. 

\textbf{ Anthropic Claude 4 Sonnet and Claude 4 Sonnet Extended:} Anthropic’s Claude models are renowned for their advanced reasoning, language comprehension, and efficient code generation. The most recent model, Claude 4 Sonnet which was released on 22 May 2025, includes an extended variant with enhanced capabilities, both of which we incorporate in our analysis.

\textbf{Google Gemini 2.5 Flash and Gemini 2.5 Pro:} Google’s strong track record in AI innovation makes its latest Gemini models essential for our benchmark. We therefore include Gemini 2.5 Flash, first released in public preview on April 17, 2025, alongside Gemini 2.5 Pro, whose public preview build launched on May 6, 2025, and is specifically promoted for advanced reasoning, mathematical skill, and code generation prowess, ensuring our evaluation is truly comprehensive.

\textbf{xAI Grok 3 - Think:} we include xAI’s Grok 3, released on February 17, 2025, and evaluate it in its dedicated “Think” mode, which is optimized for deeper reasoning, to gauge how the latest entry performs under our efficiency benchmark.

\textbf{DeepSeek DeepThink (R1): } At its launch on 20 January 2025, DeepSeek R1 (marketed as DeepThink) attracted much attention for matching the performance of costly proprietary LLMs while remaining free and requiring far fewer training resources. Due to this exceptional efficiency and cost effectiveness, we include DeepSeek DeepThink (R1) in our evaluation to measure its capabilities against the established models in our benchmark. 

% Study Approach- 
\subsection{Methodology}

% \textcolor{red}{... ensuring robust and reliable efficiency measurements. Consider adding a sentence about how correctness was verified or tested. Did you validate against known triangle counts? which i believe yes}

A critical challenge in evaluating LLMs is the risk that their outputs may reproduce memorized training data rather than demonstrate genuine reasoning. To address this, we employ a novel evaluation strategy: Instead of relying on textual prompts alone, we provide each LLM with the source files C from Bader's benchmark (described in Section~\ref{ss:data}) and instruct them to generate a functional triangle counting algorithm that integrates into the existing framework. This approach allows us to assess the models' true algorithmic reasoning capabilities, minimizing the influence of memorized solutions.

% Two-phase experimental design:

% Phase 1: Tests synthesis capability when LLMs have access to all existing implementations
% Phase 2: Tests generation capability when LLMs must create algorithms from scratch with only basic infrastructure

Our study comprises two approaches: Optimization approach and Algorithm-Synthesis approach. The optimization approach supplies each LLM with the complete collection of C source files for existing sequential triangle-counting implementations and challenges it to generate the most efficient routine. By exposing the models to every known approach, we investigated whether they can synthesize genuinely novel strategies that outperform human‑written code. The complete prompt template appears in Fig.~\ref{fig:prompt_template}.  Note that we cold-start each LLM before the prompt.

% FIGURE ENVIRONMENT
\begin{figure}[htbp] % Position preference: here, top, bottom, page
    \centering % Center the figure horizontally

    % FRAMED BOX using mdframed
    % The [roundcorner=5pt] option mimics the image's rounded corners.
    \begin{mdframed}[roundcorner=5pt]
        % Use \ttfamily for monospaced (typewriter) font inside the box
        \ttfamily
        % The content - Use \\ for line breaks and escape special LaTeX characters \{, \}, \_
        \textbf{Prompt: } The attached code implements triangle counting in a graph given in CSR format. Write a C routine tc_fast() with the API below that is the fastest sequential triangle counting code you can generate.\\[1ex]
        UINT_t tc_fast(const GRAPH_TYPE *graph)
        \\[1ex] % Add extra vertical space after this line if needed
        \textbf{Attached Files: } (bfs.c, graph.c, main.c, queue.c, tc.c)

    \end{mdframed} % End of the framed box

    % CAPTION AND LABEL
    \caption{The First Prompt Structure Template.}
    \label{fig:prompt_template} % Add a label for cross-referencing (optional)

\end{figure} % End of the figure environment

The Algorithm‑Synthesis approach evaluates how well each model can generate efficient implementations of Triangle Counting, Diameter Finding, Vertex Connectivity, Edge Connectivity, and Clique Number, when no prior code for these algorithms is provided. In this approach the provided C source contains the project’s core graph infrastructure; the algorithm implementations are deliberately absent. Each model must therefore write a new routine that integrates seamlessly with this codebase. The prompt template for this approach is shown in Fig.~\ref{fig:prompt_template2}.

After collecting the generated code for both approaches, we incorporated them into the benchmark and executed all tests to compare performance. The benchmark reports the average runtime over 10 executions for each implementation, providing statistical validity and smoothing out measurement noise to ensure robust and reliable performance measurements.

% FIGURE ENVIRONMENT
\begin{figure}[htbp] % Position preference: here, top, bottom, page
    \centering % Center the figure horizontally

    % FRAMED BOX using mdframed
    % The [roundcorner=5pt] option mimics the image's rounded corners.
    \begin{mdframed}[roundcorner=5pt]
        % Use \ttfamily for monospaced (typewriter) font inside the box
        \ttfamily
        % The content - Use \\ for line breaks and escape special LaTeX characters \{, \}, \_
        \textbf{Prompt: } The attached code implements a graph in CSR format. Write a C routine fast() with the API below that is the fastest sequential implementation for finding the \{ALGORITHM\} of a graph that can be integrated into the attached files:\\[1ex]
        UINT_t fast(const GRAPH_TYPE *graph)
        \\[1ex] % Add extra vertical space after this line if needed
        \textbf{Attached Files: } (bfs.c, graph.c, main.c, queue.c)

    \end{mdframed} % End of the framed box

    % CAPTION AND LABEL
    \caption{The Second Prompt Structure Template. \{ALGORITHM\} is replaced by "number of triangles", "diameter", "vertex connectivity", "edge connectivity", and "clique number"}
    \label{fig:prompt_template2} % Add a label for cross-referencing (optional)

\end{figure} % End of the figure environment

\vspace{0.3in}
\section{Experimental Results}

% \textcolor{red}{... algorithmic details for these two models were excluded from further analysis. It would strengthen the section to provide a brief explanation of why the incorrect models failed.}\textcolor{green}{ disagree}

% \textcolor{red}{... not groundbreaking, were effectively synthesized into workable solutions. This is insightful, consider expanding on how synthesis differs between models.}

% \textcolor{red}{... even though it did not introduce fundamentally new algorithmic concepts... Add a sentence discussing what this implies about LLM strengths and limitations (e.g., optimization over innovation).}

The experiments were conducted on Wulver, NJIT's high-performance supercomputer, using a single core of an AMD EPYC 7753 CPU @ 2.45\,GHz and 512\,GB of RAM.

\subsection{Optimization Approach}

The evaluation demonstrates that while all models examined successfully generated code that seamlessly integrated into the benchmark framework without requiring modifications, DeepSeek R1, Gemini 2.5 Flash, and Grok 3-Think produced implementations that failed to correctly count the number of triangles. Consequently, due to these incorrect outputs, the algorithmic details of the code generated by these three models were excluded from further analysis.

Most models used efficient approaches already present in the benchmark, combining optimized techniques such as degree-based vertex sorting (using the benchmark’s built-in function), hash-based intersection, the Forward Triangle Counting algorithm \cite{schank2005finding,schank2007algorithmic} and the BFS traversal \cite{bader2023triangle}. Although not groundbreaking, these strategies were effectively synthesized into workable solutions. 

Although both Gemini 2.5 Pro and Claude Sonnet 4 Extended employed similar core methods, using the Forward Counting Algorithm combined with sorting and hashing, Claude Sonnet 4 Extended achieved faster performance. This improvement indicates a more effective utilization of memory management and cache locality by the code generated from the Claude model. The code of the other models achieves their fastest running times by using higher peak memory.

ChatGPT o4-mini-high implemented the Forward Triangle Counting Algorithm for its solution. However, its implementation did not incorporate advanced optimizations such as hashing or sorting of nodes to accelerate neighbor intersection computations, leading to suboptimal time performance, particularly in graphs with high-degree vertices where efficient lookup mechanisms could reduce redundant comparisons. As a result, its implementation was third among LLM implementations, trailing behind human-written code.

ChatGPT o3 incorporated two human-written helper functions, one for graphs with more than 16,384 edges and the other for smaller instances. However, its choice of which function to apply was not optimal compared to other LLMs or the hand-written baseline, leading to a slower running time despite achieving the lowest peak memory usage compared to other LLMs.

Claude Sonnet 4 also adopted the Forward Triangle Counting Algorithm and incorporated hashing to optimize neighbor intersection operations, a strategy that theoretically should improve performance for common neighbors. However, its implementation exhibited significantly worse running time compared to other models, which shows its poor performance in implementing efficient code.  

The source code for the implementations produced by each model is available on GitHub at https://github.com/Bader-Research/LLM-triangle-counting/. Table \ref{t:results1} then summarizes their comparative performance, both execution time and maximum memory usage, on the RMAT-18 dataset \cite{chakrabarti2004r}. 

To comprehensively evaluate the performance of the generated code, we tested the corrected algorithms on graphs ranging from RMAT-6 to RMAT-18. This allowed us to observe runtime behavior across both small and large graphs. We also included an efficient human-generated implementation of Bader’s algorithm that incorporates sorting vertices, BFS traversal, and hash arrays, as well as two different implementations of the forward triangle counting algorithm with neighbor intersections based on hash.

The results in Table \ref{t:results:intel9480} show that Claude 4 Sonnet Extended - and, to a slightly lesser extent, Gemini 2.5 Pro - achieved the fastest running times by trading increased peak memory for lower execution time. Both models outperformed every other LLM and even the human‐written baseline, despite relying on existing triangle‐counting techniques rather than inventing new algorithms. This suggests that while large language models excel at optimizing and refining code, their current capabilities remain focused on improving known methods rather than developing fundamentally new algorithmic advances.

\subsection{Algorithm-Synthesis Approach}

In this approach, we evaluate the capacity of each model to produce efficient implementations of established graph algorithms without access to any pre-existing code. Performance is evaluated according to two principal criteria: Ready-To-Use (RTU) Capability and Efficiency Trade-Offs.

\subsubsection{Ready-To-Use (RTU) Capability}
% This parameter defined based on three important properties which a given code shuold have in the prespective of user.
% these properties are:
This parameter is defined based on three important properties that generated code should have from a user's perspective:
% A model is deemed RTU for a given algorithm if it generates code that meets all of the following requirements:
%the RTU is defined based on these criterias.. 

\begin{itemize}
  \item \textbf{Compilability:} The code must compile successfully without any manual modifications.
  \item \textbf{Correctness:} The code must pass all test cases in our comprehensive suite, which includes different types of graphs. 
  \item \textbf{Timeliness:} The implementation must run within an acceptable runtime threshold, ensuring practical usability.
\end{itemize}
%A model is deemed RTU for a given algorithm if it generates code that meets all of the following requirements:
If the generated code does not meet any of these criteria or even fails in a single test case, it is not considered an RTU. 

Table \ref{tab:rtu_percentage} reports the RTU rate for each model in the six target graph problems. The results indicate that Claude Sonnet 4 Extended achieves the highest RTU score, successfully generating fully compilable, correct, and timely code for 83\% of the algorithmic tasks.

\begin{table}[!t]
    
    \caption{RTU Percentage for Various Models}
    \label{tab:rtu_percentage}
    \centering
    \begin{tabular}{lc}  % <-- changed second column from 'l' to 'c'
        \toprule
        model   & RTU percentage \\
        \midrule
        Google Gemini 2.5 Flash               & 17\% \\
        Google Gemini 2.5 Pro                 & 50\% \\
        Deepseek DeepThink             & 33\% \\
        OpenAI ChatGPT o3              & 17\% \\
        OpenAI ChatGPT o4-mini-high    & 50\% \\
        xAI Grok 3 Think               & 50\% \\
        Anthropic Claude Sonnet 4                & 50\% \\
        Anthropic Claude Sonnet 4 Extended       & \textbf{83\%} \\
        \bottomrule
    \end{tabular}

\end{table}

\subsubsection{Efficiency Trade-Offs}

Beyond the status of the RTU, we also evaluate the time-memory trade-off of each model when executing the generated routines. Specifically, we record two metrics for each model-algorithm pair:
\begin{enumerate}
    \item \textbf{Relative runtime (\(t\))}—normalized so that the fastest implementation for a given algorithm has \(t = 1\).
    \item \textbf{Relative peak memory usage (\(m\))}—normalized so that the implementation that uses the least memory for the same algorithm has \(m = 1\).
\end{enumerate}

Table \ref{tab:runtime-mem} presents the relative values for the code generated by each model and its RTU algorithm. Models that did not generate a valid RTU implementation are omitted from the analysis and are indicated with a dash.

To compare the models on a single efficiency metric, we introduce a rate that combines relative runtime and peak memory consumption:
\[
\mbox{rate} = 
\begin{cases}
    1/(tm) , & \mbox{if the model produced RTU implementation}\\
    0, & \mbox{otherwise}.
\end{cases}
\]

Here, \(t\) is the code’s running time expressed as a fraction of the fastest implementation, and \(m\) is its maximum memory usage expressed as a fraction of the most memory‑efficient implementation. A higher rate therefore indicates greater overall efficiency. The overall score of each model is calculated by adding its rates in the six benchmark algorithms; The resulting totals appear in Table \ref{tab:efficiency_rates}.

\begin{table}[h]
    \caption{Efficiency rates of different models.}
    \label{tab:efficiency_rates}
    \centering
    \begin{tabular}{lc}
        \toprule
        \textbf{Model} & \textbf{Efficiency Rate} \\
        \midrule
        Google Gemini 2.5 Flash    & 0.76 \\
        Google Gemini 2.5 Pro      & 1.82 \\
        DeepSeek DeepThink     & 0.81 \\
        OpenAI ChatGPT o3           & 0.04 \\
        OpenAI ChatGPT o4-mini-high           & 1.05 \\
        xAI Grok3 Think         & 2.84 \\
        Anthropic Claude 4 Sonnet    & 2.02 \\
        Anthropic Claude 4 Sonnet Extended  & \textbf{3.11} \\
        \bottomrule
    \end{tabular}
    
\end{table}

The results indicate that Claude Sonnet 4 Extended is currently the most proficient LLM for software performance engineering of sparse graph algorithm code in terms of combined runtime and peak memory usage, achieving the highest rate (3.11). Grok 3 Think ranks second, followed by Claude Sonnet 4 in third place. 

\section{Conclusions}

% \textcolor{red}{... fundamentally novel computational strategies. Well stated, consider summarizing future directions in one sentence here to link to the next section.}

As one of the first systematic explorations of LLM-assisted software performance engineering, this empirical study establishes important foundational insights into the capabilities and potential of this rapidly evolving field. Our comprehensive evaluation of eight state-of-the-art models demonstrates that contemporary LLMs can successfully generate efficient C implementations for performance-critical graph algorithms, with all models seamlessly integrating their code into our benchmark framework during the optimization approach.
The results reveal significant efficiency differences among models, with Claude 4 Sonnet Extended and Gemini 2.5 Pro achieving the most impressive runtime and memory profiles, even outperforming human-written baselines in triangle counting. In the algorithm-synthesis approach, Claude Sonnet 4 Extended achieved the highest Ready-To-Use (RTU) rate (83\%) and generated the most efficient implementations, establishing it as the current leader for software performance engineering tasks in this domain.

Although our findings demonstrate that current LLMs excel in synthesizing, optimizing, and refining established algorithmic approaches, this represents a crucial first step in the emerging field of LLM-assisted algorithm engineering. The observed capabilities suggest that these models possess a sophisticated understanding of algorithmic principles, memory management, and performance optimization strategies. Most importantly, the potential for LLMs to eventually transcend the optimization of existing methods and invent transformative algorithmic techniques represents one of the most compelling frontiers in computational science.

This work provides essential benchmarks and methodologies for a field poised for rapid advancement, offering both practitioners and researchers the foundational tools needed to harness LLMs for high-performance computing applications while establishing the groundwork for future breakthroughs in algorithmic innovation.

\vspace{0.5in}
\section{Future Work}
%\vspace{-0.05in}

% \textcolor{red}{... graph algorithm implementation. Also mention possible directions in verifying correctness/security formally using tools like static analyzers.}

The pioneering nature of this study opens numerous avenues for advancing LLM-assisted software performance engineering, particularly as this field enters a phase of accelerated development. Several critical research directions warrant immediate exploration to unlock the transformative potential of these technologies.

\noindent
\textbf{Algorithmic Innovation and Creativity:} Future research should investigate methods to enhance the capacity of LLMs for genuine algorithmic invention, moving beyond the optimization of known techniques toward the discovery of fundamentally novel computational strategies. This includes developing prompting techniques, training methodologies, and evaluation frameworks specifically designed to foster algorithmic creativity and breakthrough thinking.

\noindent
\textbf{Massively Parallel and GPU Computing:} Extending evaluation to parallel graph algorithms and GPU architectures represents both a natural progression and a critical need, as LLMs may demonstrate superior capabilities in optimizing for complex parallel execution patterns, load balancing, synchronization, and memory hierarchy optimization in heterogeneous computing environments.

\noindent
\textbf{Theoretical Foundations and Guarantees:} Establishing rigorous theoretical frameworks for LLM-generated algorithms, including formal analyses of time complexity, memory consumption, and correctness guarantees, will be essential for deploying these techniques in high-stakes and safety-critical computational applications.

\noindent
\textbf{Domain-Specific Algorithm Engineering:} Expanding beyond graph algorithms to encompass broader algorithmic domains, including numerical methods, machine learning kernels, cryptographic implementations, and scientific computing routines, will reveal the full scope of the potential impact of LLMs on computational science.

\noindent
\textbf{Interactive and Iterative Algorithm Development:} Investigating human-AI collaboration models where LLMs work interactively with domain experts to iteratively refine and evolve algorithmic solutions, which could lead to hybrid approaches that combine human intelligence with AI optimization capabilities.

\noindent
\textbf{Real-World Deployment and Integration:} Developing frameworks to seamlessly integrate LLM-generated high-performance code into production systems, including automated testing, verification and continuous optimization pipelines that can adapt algorithms as hardware and requirements evolve.

As this field rapidly evolves, we anticipate that LLMs will increasingly demonstrate the capability to design novel approaches that fundamentally advance the state of algorithmic science, potentially revolutionizing how we approach computational problem solving across disciplines. The foundation established in this work provides a starting point for what promises to be one of the most transformative developments in algorithm engineering.

\onecolumn
% \textcolor{green}{In tables, bold best runtimes or use arrows/icons to highlight performance trends across graph sizes.}

% \textcolor{red}{In the appendix, ensure all generated code is compilable and consistently formatted.}

\begingroup
\scriptsize
\setlength{\tabcolsep}{3pt}     % default is 6pt
\renewcommand{\arraystretch}{0.9} % default is 1.0

\begin{table}
\caption{Comparative Overview of the LLMs \\ Compilable: The code compiled successfully without any modifications. Correctness: The model counted the triangles correctly. \# triangles: Numer of triangles that the model counted for RMAT18. Runtime RMAT 18: The runtime of the model on RMAT 18 in seconds. Max Mem. Us.: The relative maximum memory usage of the model on RMAT 18 by taking ChatGPT o3 as the baseline. Sort: The model sorted the vertices by degree at the start of the algorithm. Hash: The model used hash tables for intersection operations. FTC: The model used the Forward Counting algorithm in the proposed method. BFS: The model employed breadth‑first search (BFS) in its implementation. }

\centering
\begin{tabular}{lrrrrrrrrrrrrr}
 Model          & Compilable         & Correctness         & \# triangles & Runtime  & Max Mem. Us.       & Sort             & Hash            & FTC        & BFS       \\ \hline
ChatGPT o3                      & \checkmark        & \checkmark      & 101930789        & 16.814495       & 1     & \checkmark            & \checkmark               & \checkmark         & \checkmark   \\
ChatGPT o4-mini-high            & \checkmark        & \checkmark      & 101930789      & 4.714634 & 1.01      & \texttimes      & \texttimes      & \checkmark  & \texttimes \\
Claude 4 Sonnet               & \checkmark       & \checkmark      & 101930789     & 192.777913 & 1     & \texttimes      & \checkmark      & \checkmark  & \texttimes  \\
Claude 4 Sonnet Extended      & \checkmark       & \checkmark      & 101930789    & \textbf{0.624521}   & 1.45    & \checkmark      & \checkmark      & \checkmark  & \texttimes \\
Gemini 2.5 Pro                & \checkmark       & \checkmark      & 101930789   & 0.665562    & 1.46    & \checkmark      & \checkmark      & \checkmark  & \texttimes   \\
Gemini 2.5 Flash                & \checkmark       & \texttimes      & 305792367  & NA   & NA      & NA      & NA      & NA  & NA   \\

DeepSeek DeepThink (R1)         & \checkmark      & \texttimes     & 203861578    & NA  & NA    & NA     & NA      & NA  & NA   \\
Grok 3-Think                          & \checkmark      & \texttimes     & 3812543  & NA  & NA      & NA      & NA      & NA  & NA   \\[12pt]

\end{tabular}
\label{t:results1}
\end{table}

\begin{table}
\caption{Execution time of each model (in seconds). \\ \# Vertices: number of vertices in the graph. \# Edges: number of edges in the graph. \# triangles: Number of Triangles in the graph. o3: OpenAI ChatGPT o3. o4MH: OpenAI ChatGPT o4-mini-high. Cl 4: Anthropic Claude Sonnet 4. Cl 4 Ex.: Anthropic Claude Sonnet 4 Extended. Gemi.P: Gemini 2.5 Pro. BaderBFS: Bader Algorithm based on BFS. %tc_bader_forward_hash_degOrd
Forward 1: First implementation based on Forward Triangle Counting.%tc_forward_hash_skip
Forward 2: Second implementation based on Forward Triangle Counting.%tc_forward_hash_degOrd
}

\centering
\begin{tabular}{lrrrrrrrrrrrrr}
 Graph          & \# Vertices         & \# Edges         & \# triangles & o3   & o4MH         & Cl 4      & Cl 4 Ex.           & Gemi.P          & BaderBFS  &  Forward 1 & Forward 2     \\ \hline

RMAT 6          & 64        & 1024      & 9100      & \textbf{0.000022} & 0.000028 & 0.000230 & 0.000033 & 0.000023 & 0.000029 & \textbf{0.000022} & 0.000023 \\	
RMAT 7          & 128       & 2048      & 18855     & 0.000059 & 0.000125 & 0.000303 & \textbf{0.000050} & 0.000060 & 0.000070 & 0.000053 & 0.000056 \\	
RMAT 8          & 256       & 4096      & 39602     & 0.000149 & 0.000371 & 0.001019 & \textbf{0.000124} & 0.000150 & 0.000204 & 0.000155 & 0.000154 \\
RMAT 9          & 512       & 8192      & 86470     & 0.000419 & 0.000992 & 0.003197 & \textbf{0.000320} & 0.000360 & 0.000488 & 0.000374 & 0.000386 \\
RMAT 10         & 1024      & 16384     & 187855    & 0.000990 & 0.002567 & 0.010182 & \textbf{0.000727} & 0.000831 & 0.001126 & 0.000891 & 0.000859 \\
RMAT 11         & 2048      & 32768     & 408876    & 0.002312 & 0.006473 & 0.034632 & \textbf{0.001707} & 0.001918 & 0.002364 & 0.002048 & 0.001882  \\
RMAT 12         & 4096      & 65536     & 896224    & 0.005334 & 0.017888 & 0.115028 & \textbf{0.003860} & 0.004316 & 0.005302 & 0.004758 & 0.004389 \\	
RMAT 13         & 8192      & 131072    & 1988410   & 0.012643 & 0.045412 & 0.408823 & \textbf{0.009144} & 0.010036 & 0.011355 & 0.011520 & 0.010052  \\	
RMAT 14         & 16384     & 262144    & 4355418   & 0.031400 & 0.115286 & 1.364177 & \textbf{0.022311} & 0.023743 & 0.026098 & 0.029162 & 0.023728  \\	
RMAT 15         & 32768     & 524288    & 9576800   & 1.094913 & 0.300161 & 4.770884 & \textbf{0.053767} & 0.056490 & 0.058261 & 0.076366 & 0.057285 \\	
RMAT 16         & 65536     & 1048576   & 21133772  & 2.727890 & 0.756705 & 16.158756 & \textbf{0.128004} & 0.134363 & 0.131487 & 0.189328 & 0.135958  \\	
RMAT 17         & 131072    & 2097152   & 46439638  & 6.737418 & 1.860513 & 56.074888 & \textbf{0.283528} & 0.298584 & 0.307493 & 0.465803 & 0.303541 \\	 
RMAT 18         & 262144    & 4194304   & 101930789 & 16.814495 & 4.714634 & 192.777913 & \textbf{0.624521} & 0.665562 & 0.723090 & 1.259261 & 0.733882   \\	 
[12pt]

\end{tabular}
\label{t:results:intel9480}
\end{table}

\begin{table}[ht]
\caption{Relative running times (Time) and relative peak memory usage (Mem) for each generated code; the numeric suffix appended to an algorithm’s name denotes the RMAT graph scale on which it was evaluated.}
  \label{tab:runtime-mem}
  \centering
  %— makes the grid a bit roomier; adjust or delete to taste
  %\renewcommand{\arraystretch}{1.25}

  % Wrap in adjustbox so it never spills past the page margins
  \begin{adjustbox}{center}
  
  \begin{tabular}{|l|c|c|c|c|c|c|c|c|c|c|c|c|}
    \hline
    % ---------- TOP HEADER (spans two columns per algorithm) ----------
    \multicolumn{1}{|c|}{} &
    \multicolumn{2}{c|}{\shortstack{\textbf{Triangle Counting}\\\textbf{15}}} &
    \multicolumn{2}{c|}{\shortstack{\textbf{Diameter Finding}\\\textbf{15}}} &
\multicolumn{2}{c|}{\shortstack{\textbf{Vertex Connectivity}\\\textbf{15}}} &
\multicolumn{2}{c|}{\shortstack{\textbf{Edge Connectivity}\\\textbf{12}}} &
\multicolumn{2}{c|}{\shortstack{\textbf{Clique Number}\\\textbf{12}}} &
\multicolumn{2}{c|}{\shortstack{\textbf{Chromatic Number}\\\textbf{15}}} \\ 
    \hline
    % ---------- SUB‑HEADER (repeats Time / Mem) ----------
    & \textbf{R. Time} & \textbf{R. Mem} &
      \textbf{R. Time} & \textbf{R. Mem} &
      \textbf{R. Time} & \textbf{R. Mem} &
      \textbf{R. Time} & \textbf{R. Mem} &
      \textbf{R. Time} & \textbf{R. Mem} &
      \textbf{R. Time} & \textbf{R. Mem} \\
    \hline
    % ---------- DATA ROWS (add/duplicate as needed) ----------
    Google Gemini 2.5 Flash &-        &-    &1.32 &1 &-         &- &-         &-     &-         &- &- &- \\ \hline
    Google Gemini 2.5 Pro   &20.68 &1    &1.30 &1 &-         &- &1 &1     &-         &- &- &- \\ \hline
    DeepSeek DeepThink      &37.96 &1    &1.27 &1 &-         &- &-         &-     &-         &- &- &- \\ \hline
    OpenAI ChatGPT o3       &-        &-    &-         &- &-         &- &1.64 &15.85 &-&- &- &- \\ \hline
    \shortstack{OpenAI ChatGPT\\o4-mini-high} 
                            &4.78 &1    &1.28 &1 &-         &- &1.05 &15.85 &-         &- &- &- \\ \hline
    xAI Grok3 Think         &1 &1    &1.19 &1 &-         &- &-         &-     &1  &1 &- &- \\ \hline
    Anthropic Claude 4      &57.93 &1.04    &1 &1 &- &- &-         &-     &-         &- &1 &1 \\ \hline
    \shortstack{Anthropic Claude 4\\Extended} 
                            &8.60 &1.04 &1 &1 &1  &1 &12.05 &31.51 &-         &- &1 &1 \\ \hline
  \end{tabular}
  \end{adjustbox}

\end{table}

\twocolumn

%\vspace{-0.05in}
%\clearpage

\bibliographystyle{IEEEtran}
\bibliography{ref}

\end{document}